\documentclass[11pt,twoside]{article}
\usepackage{./asp2014}
\usepackage{color}

\aspSuppressVolSlug
\resetcounters

\begin{document}

\newcommand\todo[1]{\textcolor{blue}{\textbf{#1}}}
\title{Magnetic fields in forming stars with the ngVLA}
\author{
Charles L. H. Hull,$^{1,2,3}$
Carlos Carrasco-Gonz\'alez,$^4$
Peter K. G. Williams,$^5$
Josep M. Girart,$^6$
Timothy Robishaw,$^7$
Roberto Galv\'an-Madrid,$^4$
and Tyler Bourke$^8$ \\
\affil{$^1$National Astronomical Observatory of Japan, NAOJ Chile Observatory, Alonso de C\'ordova 3788, Office 61B, Vitacura 763 0422, Santiago, Chile; \email{chat.hull@nao.ac.jp}}
\affil{$^2$Joint ALMA Observatory, Alonso de C\'ordova 3107, Vitacura 763 0355, Santiago, Chile}
\affil{$^3$NAOJ Fellow}
\affil{$^4$Instituto de Radioastronom\'ia y Astrof\'isica (IRyA-UNAM), Universidad Nacional Aut\'onoma de M\'exico, Campus Morelia, Apartado Postal 3-72, 58090 Morelia, Michoac\'an, Mexico}
\affil{$^5$ Harvard-Smithsonian Center for Astrophysics, 60 Garden St, MS 20, Cambridge, MA 02138, USA}
\affil{$^6$Institut de Ci\`encies de l'Espai (CSIC-IEEC), Campus UAB, Carrer de Can Magrans S/N, E-08193 Cerdanyola del Vall\`es, Catalonia, Spain}
\affil{$^7$Dominion Radio Astrophysical Observatory, National Research Council Canada, PO Box 248, Penticton, BC, V2A 6J9, Canada; }
\affil{$^8$Square Kilometre Array Organization, Jodrell Bank Observatory, Lower Withington, Macclesfield, Cheshire SK11 9DL, UK;}
}

\paperauthor{Charles L. H. Hull}{chat.hull@nao.ac.jp}{0000-0002-8975-7573}{National Astronomical Observatory of Japan}{NAOJ Chile Observatory}{Vitacura}{Santiago}{}{Chile}
\paperauthor{Carlos Carr\'asco-Gonz\'alez}{c.carrasco@irya.unam.mx}{0000-0003-2862-5363}{Instituto de Radioastronom\'ia y Astrof\'isica (IRyA)}{Universidad Nacional Aut\'onoma de M\'exico (UNAM)}{}{}{Morelia}{Mexico}
\paperauthor{Peter K. G. Williams}{pwilliams@cfa.harvard.edu}{0000-0003-3734-3587}{Harvard-Smithsonian Center for Astrophysics}{}{Cambridge}{MA}{02138}{USA}
\paperauthor{Josep M. Girart}{girart@ice.cat}{0000-0002-3829-5591}{Institut de Ci\`encies de l'Espai}{}{Cerdanyola del Vall\`es}{Catalonia}{E-08193}{Spain}
\paperauthor{Timothy Robishaw}{tim.robishaw@nrc-cnrc.gc.ca}{0000-0002-4217-5138}{National Research Council Canada}{Dominion Radio Astrophysical Observatory}{Penticton}{BC}{V2A 6J9}{Canada}
\paperauthor{Roberto Galv\'an-Madrid}{r.galvan@irya.unam.mx}{0000-0003-1480-4643}{Instituto de Radioastronom\'ia y Astrof\'isica}{UNAM Campus Morelia}{}{Morelia}{Apdo. Postal 72-3 (Xangari)}{M\'exico}
\paperauthor{Tyler Bourke}{T.Bourke@skatelescope.org}{ORCID_Or_Blank}{Square Kilometre Array Organisation}{Jodrell Bank Observatory}{Macclesfield}{Cheshire}{SK11 9DL}{UK}

\section{Introduction}

The magnetic field plays an important role in every stage of the star-formation process from the collapse of the initial protostellar core to the star's arrival on the main sequence.  Consequently, the goal of this science case is to explore a wide range of magnetic phenomena that can be investigated using the polarization capabilities of the Next Generation Very Large Array (ngVLA).  These include (1) magnetic fields in protostellar cores via polarized emission from aligned dust grains, including in regions optically thick at wavelengths observable by the Atacama Large Millimeter/submillimeter Array (ALMA); (2) magnetic fields in both protostellar cores and molecular outflows via spectral-line polarization from the Zeeman and Goldreich-Kylafis effects; (3) magnetic fields in protostellar jets via polarized synchrotron emission; and (4) gyrosynchrotron emission from magnetospheres around low-mass stars.

\section{Scientific Importance \& Anticipated Results}

\subsection{Dust polarization in low-mass protostellar objects}

Mapping the morphology of magnetic fields in the cores and envelopes surrounding young, low-mass protostars is critical to further our understanding of how magnetic fields affect the star-formation process at early times.  Under most circumstances, and for the typical grain sizes found in the ISM ($\ll$100~$\mu$m), elongated dust grains are thought to align themselves with their long axes perpendicular to magnetic field lines \citep[e.g.,][]{Hildebrand1988, Lazarian2007, Hoang2009, Andersson2015}, resulting in thermal radiation from the grains that is polarized perpendicular to the magnetic field (see \citealt{Heiles1993} for a description of not only linear dust polarization, but also the other polarization mechanisms that are mentioned in the following sections).  Thus, this type of observation can be used to infer the magnetic field morphology in the plane of the sky.  

\begin{figure*}
\begin{center}
\includegraphics[width=0.7\textwidth, clip, trim=0cm 0cm 0cm 0cm]{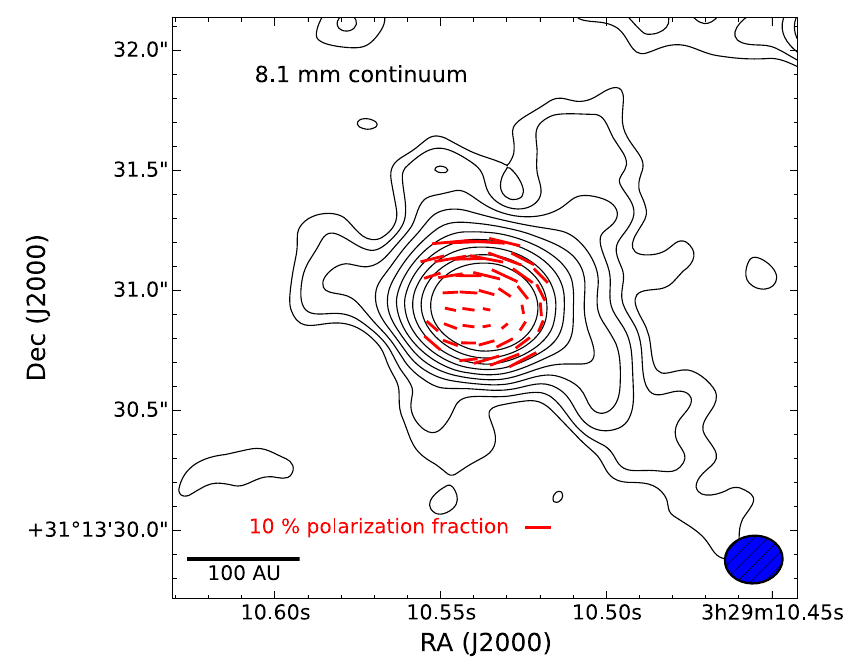}
\end{center}
\vspace{-1em}
\caption{\small
VLA 8\,mm polarization observations of the embedded Class 0 protostellar source NGC 1333-IRAS 4A \citep{Cox2018}.  Contours show total intensity (Stokes $I$) emission; red line segments show the inferred magnetic field, rotated by 90$^\circ$ relative to the dust polarization.  IRAS 4A is one of the brightest low-mass protostellar sources known; typical sources are not bright or polarized enough for dust polarization to be detected toward them with the JVLA.
}
\label{fig:iras4a}
\end{figure*}

Polarized thermal dust emission from both low- and high-mass protostellar objects has been studied extensively at millimeter wavelengths at 1--2$\arcsec$ resolution by the Berkeley Illinois Maryland Association millimeter array \citep[BIMA; e.g.,][]{Rao1998, Girart1999, Lai2001, Matthews2005, Cortes2006a, Kwon2006}, the Submillimeter Array \citep[SMA; e.g.,][]{Girart2006, Girart2009, Tang2009a, Alves2011, Zhang2014}, the Combined Array for Research in Millimeter-wave Astronomy \citep[CARMA; e.g.,][]{Hull2013, Hughes2013, Hull2014, Stephens2014, SeguraCox2015}, and ALMA \citep[e.g.,][]{Cortes2016, Hull2017a, Hull2017b, Cox2018, Maury2018}.  With the upgraded Karl G. Jansky Very Large Array (JVLA), it has recently become possible to begin performing dust polarization studies at centimeter wavelengths, using the highest-frequency bands available: K, Ka, and Q bands (18--50\,GHz).  Both \citet[][see Figure \ref{fig:iras4a}]{Cox2015} and \citet{Liu2016} studied the polarized dust emission at 37 and 43\,GHz from NGC 1333-IRAS 4A, one of the first sources to be imaged in full polarization and high resolution at submillimeter wavelengths using the SMA \citep{Girart2006}.  More recently, \citet{Liu2018} have studied the 40--48\,GHz polarization toward the Class 0 protostar IRAS 16293, another bright source that was an early target for millimeter-wave polarization observations \citep{Rao2009, Rao2014}.

For the JVLA, NGC 1333-IRAS 4A and IRAS 16293 are two of the very few sources with polarized dust emission bright enough that they can be imaged within a few hours.  A typical protostellar core may be $\sim$\,10\,$\times$ fainter than IRAS 4A or IRAS 16293.  In order to probe the magnetic field in the innermost $\sim$\,1000--100\,au regions (which can be resolved with the 1000--10\,mas scales recoverable with the ngVLA) of these more ``normal'' objects, we need the sensitivity and the resolution of the ngVLA at the $\sim$\,30--90\,GHz wavelengths where there is still appreciable dust emission, but where material is still optically thin even in the densest inner regions.  For example, the peak flux of IRAS 4A at 30\,GHz as detected by \citet{Cox2015} is $\sim$\,3000\,$\mu$Jy in an 0.25$\arcsec$ synthesized beam.  In <\,1\,hr of integration time, the ngVLA will be able to detect 1\% polarized dust emission at a frequency of 30\,GHz at a level >\,3$\sigma$ in objects 10\,$\times$ fainter than IRAS 4A.

\subsubsection{Polarization in protoplanetary disks}

Regarding protoplanetary disk polarization at long wavelengths: a notable non-detection is HL Tau---one of the brightest known protoplanetary disks---where exceptionally deep 7\,mm VLA observations by \citet{CarrascoGonzalez2016} were unable to detect dust polarization, despite having an rms noise of $\sim$\,3\,$\mu$Jy\,beam$^{-1}$ (after integrating for nearly 15\,hrs on source), simply because the dust emission is too faint at those long wavelengths.  Polarization fractions of $\sim$\,1\% are expected in protostellar disks; thus, an order-of-magnitude improvement in sensitivity is needed in order to detect polarization in these sources.


HL Tau is $\sim$\,10\,$\times$ brighter than a ``typical'' extended protoplanetary disk: for example, the subsample of transition disks in the ALMA survey of $\rho$ Oph by \citet{Cox2017} has a median integrated flux of $\sim$\,260\,mJy at 850\,$\mu$m, whereas HL Tau has an integrated flux of brightness of $\sim$\,2.4\,Jy at the same wavelength \citep{Stephens2017b}.  Assuming this approximate factor of 10 ratio holds at longer wavelengths, a typical large disk would have a peak 3\,mm flux of $\sim$\,3\,mJy in a $\sim$0.5$\arcsec$ beam, 10\,$\times$ lower than the value reported in \citet{Stephens2017b}.  In <\,1\,hr of integration time, the ngVLA will be able to detect 1\% polarized dust emission at a wavelength of 3\,mm at a level >\,3$\sigma$ in such an object, enabling studies of long-wavelength polarization in a statistically significant sample of many of the famous, extended disks present in a wide range of star-forming regions.
These studies will also allow us to characterize the polarization in regions optically thick at ALMA wavelengths \citep[e.g., in HL Tau;][]{ALMA2015, CarrascoGonzalez2016, Jin2016}; will enable us to characterize the polarization properties of grains by observing the polarization spectrum---i.e., the source polarization as a function of wavelength---from infrared to (sub)millimeter to centimeter wavelengths; and will allow us to better understand the physics behind the transition from magnetic alignment of dust grains to other polarization mechanisms (see below), which are thought to be dependent on dust-grain growth.

Recent ALMA polarization observations toward several protoplanetary disks have revealed different mechanisms causing the polarization: self-scattering at 870\,$\micron$ (observational: \citealt{Kataoka2016b, Stephens2017b, CFLee2018, Girart2018, Hull2018a}; theoretical: \citealt{Kataoka2015, Yang2017}), alignment with the dust-emission gradient at 3\,mm (observational: \citealt{Kataoka2017}; theoretical: \citealt{Tazaki2017}), and a seemingly linear combination of the two at 1.3\,mm \citep{Stephens2017b}.  Recent (sub)millimeter ALMA polarization observations of younger, more embedded Class 0/I protostars have even shown hints of scattering in the innermost, $\sim$\,100\,au regions of a few sources \citep{Cox2018, Harris2018, Sadavoy2018}.  However, self-scattering is a highly wavelength-dependent phenomenon, and it is maximum around a wavelength of $2 \pi a_{\rm max}$, where $a_{\rm max}$ is the maximum grain size. Thus, at long wavelengths far from this value (i.e., those accessible to the ngVLA), it may be possible that grains aligned with the magnetic field dominate the dust polarization.

Considering that ALMA will eventually have Band 1 ($\sim$\,40\,GHz), and that it already has a functioning polarization system at Band 3 ($\sim$\,100\,GHz), lower-frequency K and Ka band (18--40\,GHz) observations with the ngVLA would be complementary, and possibly unique to the instrument.  Also, without considering the unknown polarization efficiency as a function of wavelength, 18--40\,GHz observations are easier than higher Q band (40--50\,GHz) observations at the JVLA site due to the weather constraints.

\subsection{Spectral-line polarization: Zeeman and Goldreich-Kylafis effects}


Molecular and atomic lines are sensitive to magnetic fields, which cause their spectral levels to split into magnetic sub-levels. When threaded by a magnetic field, atomic hydrogen and molecules with a strong magnetic dipole moment will have the degeneracy in magnetic sub-levels lifted for states with non-zero angular momentum.  Examples lying within the nominal 1.2--116\,GHz frequency range of the ngVLA include lines from H at 1.42\,GHz; OH at 1.61--1.72\,GHz \& 6.01--6.05\,GHz; CH$_3$OH at 6.7, 36, \& 44\,GHz; C$_4$H at 9.5 \& 19.0\,GHz; CCS at 11, 22, 33, \& 45\,GHz; C$_2$H at 87\,GHz; SO at 13, 30, 63, 86, \& 99\,GHz; and CN at 113.1--113.6\,GHz.  This lifting of the degeneracy will split the radio frequency transitions into a number of linearly and elliptically polarized components slightly separated in frequency.  This is known as the Zeeman effect \citep{Troland1986, Heiles1993, Crutcher1999}.  Measuring this effect is the only way to directly measure the magnetic field strength (yielding either the line-of-sight field strength for Zeeman-broadened transitions, or the entire field strength for transitions that are completely split, e.g., Galactic OH masers\footnote{\,Linear and circular polarization from maser emission of centimeter and millimeter lines of SiO, CH$_3$OH, H$_2$O, and OH would allow us to probe the dense environment around young stellar objects and the circumstellar material around evolved stars.}).  Directly measuring the magnetic field strength is essential to achieve a more robust understanding of the dynamical relevance of the magnetic fields before and during the collapse of the star-forming core.  Galactic OH masers can also be used to map out the large-scale magnetic field in the Milky Way since the star-forming regions in which they're found seem to retain information about the large-scale field in their vicinity \citep{Davies1974,Reid1990}.  The ngVLA will play a critical role in extending Zeeman measurements beyond our Galaxy by detecting magnetic fields from extragalactic OH masers in nearby galaxies like M33 and M31 and from OH megamasers in distant starbursting galaxies \citep{Robishaw2008}.

For other types of molecules, linear polarization can arise whenever an anisotropy in the radiation field yields a non-LTE population of magnetic sub-levels. This is the so-called Goldreich-Kylafis (G-K) effect, which traces the plane-of-sky magnetic field orientation \citep{Goldreich1981, Goldreich1982, Kylafis1983, Deguchi1984}. This effect is expected to be easiest to observe where the lines have an optical depth of $\sim$\,1; when the ratio of the collision rate to the radiative transition rate (i.e., the spontaneous emission rate) is also $\sim$\,1; and for the lowest rotational transitions of simple molecules such as CO, CS, HCN, SiO, or HCO$^+$. So far, the G-K effect has been measured mostly around molecular outflows \citep[e.g.,][]{Girart1999, Lai2003, Cortes2005, Vlemmings2012, Girart2012, Ching2016}. This suggests that G-K observations with the ngVLA at 7\,mm (SiO, CS) or 3\,mm (CO, SiO, HCO$^+$) will provide new insight into the properties of the putative helical magnetic field near the launching point of bipolar outflows around young stellar objects, where a significant fraction of the outflow appears to be molecular.  Molecular outflows from protostellar objects tend to have a parabolic, multi-layered, ``onion-skin'' type structure, with the lowest/highest-velocity material located in the outermost/innermost shells. Therefore, measuring the polarization at different velocities enables us to reconstruct the 3-dimensional magnetic field morphology in the source.

Polarization from thermal lines requires detection of emission with a high signal-to-noise ratio at high spectral resolution.  The layout of the ngVLA (including the high-density central core of antennas) will yield excellent brightness sensitivity over a range of spatial scales relevant to star formation (approximately 10--1000\,mas).  The types of sources where the ngVLA would have a clear science impact detecting the Zeeman splitting and the G-K effect include the inner regions of various types of star-forming cores (e.g., warm cores, hot corinos, hot cores), proto/circumstellar envelopes, the launching regions of molecular outflows and jets, and disks.

\subsection{Synchrotron emission from protostellar jets}

Jets play an important role in the star-formation process. They are believed to regulate accretion from the disk onto the protostar at the earliest (Class 0/I) stages of star formation. However, jets are not only important in the field of the star formation---protostellar jets are usually considered a less energetic manifestation of the powerful relativistic jets driven by supermassive black holes at the center of active galaxies or stellar-mass black holes in our own Galaxy \citep[e.g.,][]{deGouveiadalPino2005}. Despite their importance, we still do not know how protostellar jets are launched and collimated. In the case of relativistic jets, observations of polarized synchrotron emission suggest that a helical configuration of the magnetic field is responsible for the collimation of the jet \citep[e.g.,][]{Lyutikov2005, JLGomez2008}. In the case of protostellar jets, theoretical models also point to a fundamental role of the magnetic field, which is thought to have a helical configuration \citep[e.g.,][]{Shu1994}; helical magnetic fields are a consequence of the rotation of the protostar-disk system. As for the collimation of jets, they may be collimated internally. However, other collimation mechanisms by external agents, such as ordered magnetic fields present in the protostar's neighborhood, have also been proposed \citep[e.g.,][]{Albertazzi2014}; some observations are consistent with external collimation mechanisms \citep[e.g.,][]{CarrascoGonzalez2015}.

\begin{figure}
\begin{center}
\includegraphics[width=0.9\textwidth, clip, trim=0cm 0cm 0cm 0cm]{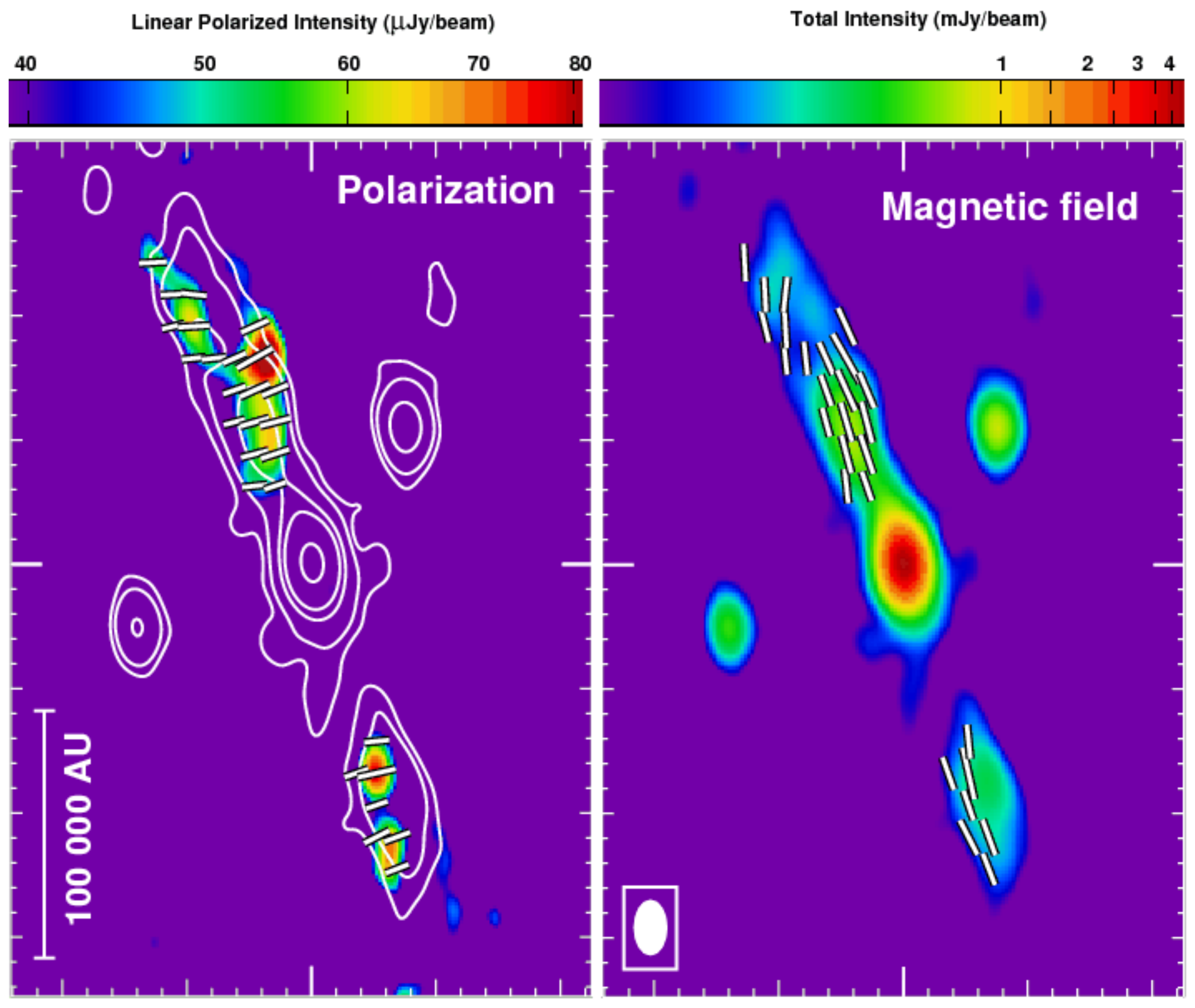}
\end{center}
\vspace{-1.5em}
\caption{\small
Observations of linearly polarized emission at 6\,cm in the protostellar jet HH 80-81 \citep{CarrascoGonzalez2010}. This is the first and only time that polarized emission has been detected in a protostellar jet. This result demonstrated that detecting polarized synchrotron emission is possible in protostellar jets, and allowed the study of the magnetic field in this object using techniques similar to those used in relativistic jets. However, the weakness of the emission and the likelihood of large rotation measures make it extremely difficult to extend this kind of study to other wavelengths and/or protostellar jets using current instruments. \emph{Left:} Color scale is the linearly polarized emission at 6\,cm. Line segments indicate the polarization orientation of the electric field. Contours are the total 6\,cm emission from the protostellar jet. \emph{Right:} White line segments indicate the orientation of the inferred (projected on the plane of the sky) magnetic field obtained from the polarization measurements. Color scale is the total 6 cm emission of the jet. The apparent magnetic field is parallel to the direction of the jet. This is consistent with a helical magnetic field configuration. However, in order to to infer the 3-dimensional configuration of the magnetic field it is necessary to study the polarization properties at several wavelengths.
}
\label{fig:hh8081}
\end{figure}

The magnetic field is extremely difficult to study in the case of protostellar jets. Jets have been observed in great detail at radio wavelengths, although they emit mainly continuum thermal radiation \citep[e.g.,][]{Anglada1996}, which contains no information about the magnetic field. However, in the last decades, very sensitive observations have shown that synchrotron emission can also be present in some protostellar jets (see \citealt{CarrascoGonzalez2010} and references therein). While the material in the jet moves at velocities of a few $\times$ 100\,km\,s$^{-1}$, it has been shown that under some conditions, a population of particles can be accelerated up to relativistic velocities in strong shocks against the ambient medium \citep[e.g.,][]{RodriguezKamenetzky2016}. Under these conditions, these accelerated particles can emit synchrotron emission, thus allowing the study of the magnetic field in the jet via synchrotron polarization. However, synchrotron emission in these objects is extremely weak, with intensities on the order of only a few $\times$ 10\,$\mu$Jy in Stokes $I$. At the beginning of the 2000s, we only knew of a handful of bright objects that show hints of synchrotron emission (i.e., negative spectral indices at centimeter wavelengths). In the last few years, more candidates have been detected as a result of the recently improved sensitivity in the ATCA and JVLA \citep[e.g.,][]{Purser2016, Osorio2017}. Yet, it has only been possible to detect linearly polarized emission in a single object, HH 80-81 \citep[][see Figure \ref{fig:hh8081}]{CarrascoGonzalez2010}. Even though HH 80-81 is the brightest known protostellar jet, this unique detection at a single wavelength (6\,cm) was only possible with significant observational effort (15\,hrs of observations with the pre-upgrade VLA). Therefore, given the sensitivity of current instruments, it is still prohibitive to extend this study to multiple wavelengths in a sample of several protostellar jets.

In order to infer the 3-dimensional configuration of the magnetic field in the jet it is necessary to study the properties of the polarization of the synchrotron emission at several wavelengths. For example, for a helical magnetic field, we expect to observe gradients of the polarization fraction and the rotation measure (RM) across the jet width \citep{Lyutikov2005}. This implies the necessity to observe with high sensitivity to detect very low polarization fractions in intrinsically weak emission, and with high angular resolution to resolve the jet width (a few $\times$ 10\,milliarcseconds). In principle, these requirements could be reached by the JVLA. However, protostellar jets are very dense (electron densities of the order of a few $\times$ 1000\,cm$^{-3}$), and thus large RMs are expected. High sensitivity is currently achieved by averaging large frequency ranges. However, this introduces depolarization across the bandwidth when large RMs are present, rendering the polarized signal impossible to detect. It is therefore necessary to obtain high sensitivities even in narrow frequency ranges, which can only be achieved by increasing the observation time and/or the number of antennas in the interferometer. 

Finally, it is not only desirable to detect polarization at several frequencies, but also at very different physical scales. The jet is expected to show complex morphology: extended diffuse emission, bow shocks, extended lobes, highly collimated parts, and compact internal shocks \citep[e.g.,][]{RodriguezKamenetzky2017}. We also expect to see changes in the density across and along the jet. All of these characteristics imply dramatic changes of the properties of the emission (e.g., RM, brightness, and polarization fraction) at different scales. These changes can actually be used to obtain information about the jet's physical structure by, for example, using new wide-band spectropolarimetry modeling techniques that have been successfully applied to very bright, but unresolved, radio galaxies \citep[e.g.,][]{OSullivan2017, Pasetto2018}. While this type of study would be very difficult to perform in distant extragalactic jets, in nearby protostellar jets it could be performed simply by being sensitive simultaneously to scales from a few milliarcseconds to several arcseconds. This will be easily achieved by an interferometer such as the ngVLA, with a large number of antennas spread over a few thousand kilometers.

\subsection{Gyrosynchrotron from active magnetospheres around young stellar objects}

Gyrosynchrotron radio emission from active magnetospheres is present in
low-mass young stellar objects (YSOs) \citep[e.g.,][]{Andre1996} as well as their more evolved
counterparts \citep{Gudel1993}, including our Sun \citep{Nita2004}. Studying
this emission in pre-main-sequence stars sheds light on their magnetic
activity and the overall ``space weather'' that they drive, which has
important effects on the chemistry and evolution of the surrounding
protoplanetary and debris disks \citep[e.g.,][]{Wilner2000}.

The characterization of magnetospheric radio emission is also necessary for the
proper analysis of YSO dust emission, because the two processes lead to
emission at overlapping radio frequencies---in particular, the range of
frequencies where the ngVLA will be most powerful. Non-thermal
gyrosynchrotron emission can be distinguished from thermal emission mainly
through its variability on timescales from minutes to months
\citep{Forbrich2015}, its generally (but not always) negative spectral index
\citep{Garay1996}, circular polarization fraction of a few to $\sim$20\%
\citep{Andre1996}, and its brightness temperatures $\gg${}$10^4$\,K
\citep{OrtizLeon2017a}. Even the upgraded JVLA, however, has difficulty
achieving the performance needed to fully understand YSO radio emission at centimeter
wavelengths. In a study using the JVLA, \citet{Liu2014}
characterized the radio flux, variability, and circular polarization of YSOs
from Class 0 to Class III, achieving a noise of $\sim$20\,$\mu$Jy using
observing blocks of a few hours and averaging over a few GHz of bandwidth. The
results were consistent with earlier findings
\citep[e.g.,][]{Gibb1999} showing that centimeter radio emission from YSOs (1) is
dominated by free-free emission from collimated winds or radio jets in the early Class
0 and I stages \citep{Anglada2015}; (2) is difficult to detect in Class
IIs \citep{GalvanMadrid2014}; and (3) is detectable in some but not all
Class III YSOs \citep{Forbrich2007, Dzib2013}. The emission from Class IIIs appears
to be dominated by gyrosynchrotron radiation, but in Class IIs there are
individual case studies pointing toward either free-free or gyrosynchrotron
\citep{Macias2016, GalvanMadrid2014}.

The high sensitivity, wide frequency coverage, and angular resolution of the
ngVLA would transform studies of both thermal and non-thermal YSO radio
emission, allowing robust disentanglement of the two processes in snapshot
observations as short as a few seconds. Scaling the nominal sensitivities
reported by \citet{Carilli2015}, the ngVLA could obtain full-polarization
measurements with noise levels of $\sim$10~$\mu$Jy in 5-second integrations.
Moreover, the large simultaneous bandwidth would allow full characterization
of the non-thermal radio SEDs and their overlap with free-free or dust thermal
emission truly simultaneously, something that has not been possible with the
VLA. It is quite likely that the non-thermal SEDs of YSOs have complexities
due to multiple magnetospheric components, binary interactions
\citep{Salter2010}, and other phenomena yet to be discovered.

The same attributes of the ngVLA that will make it a transformative tool for
studies of YSO magnetospheres will produce similar gains in studies focusing
on other objects with comparable magnetic properties. In particular, the
magnetospheres of fully-convective YSOs can be dipolar and rotation-dominated,
with strong commonalities to those found in very-low-mass stars and brown
dwarfs, magnetic chemically peculiar (MCP) stars, and even the Solar System
giant planets \citep{Donati2009, Schrijver2009}. The magnetism of these
objects is quite different from that of the Sun, being driven by a
fully-convective dynamo and resulting in large-scale current systems, coherent
maser emission, and potentially trapped particle radiation belts
\citep{Hallinan2006, Williams2017}. The resulting radio emission is generally
quite broadband and highly polarized \citep[e.g.,][]{Williams2015}, and
current studies are often sensitivity-limited; for example, the JVLA
is unable to detect even the 2\,pc-distant ultracool dwarf binary Luhman~16AB
\citep{Osten2015}. The ngVLA will therefore enable an entirely new class of
time-resolved, full-polarization studies of magnetospheric activity in YSOs
and beyond.

\section{Measurements Required}

\vspace{-0.1in}
Continuum polarization observations are relatively straightforward, generally requiring a single-pointing per source, integrating down to the point where a $\sim$\,1--3\% polarization signal can be detected at the 3+ sigma confidence level across the desired region of the source.  Spectral line polarization requires high spectral resolution, where the rule of thumb is $\sim$\,6 channels across the full-width-at-half-max (FWHM) of the line. This is especially important for Zeeman observations of masers, where a velocity resolution of <\,0.1 km/s is required. Polarization levels can be low, on the order of a few $\times$ 0.1\%, so an accurate characterization of the instrumental polarization is required, down to a level of $\sim$\,0.1\%, similar to what is achievable with the ALMA polarization system.  The more compact sources (disks, protostellar envelopes, etc.) will generally fill only a fraction of the primary beam; however, for more extended sources, and in the event that there is more than one source (e.g., a few disks or protostars) in one field of view, either separate pointings and/or mosaics will have to be performed, or the wide-field polarization characteristics will have to be well characterized in order to correct for off-axis effects (see, e.g., work done by \citealt{Jagannathan2017, Jagannathan2018}). Other standard polarization-specific calibration requirements involve solving for leakages ($D$-terms) and the $XY$ (or $RL$) phase of the reference antenna.

The design of the ngVLA antennas and feeds should be informed by careful
consideration of polarimetric goals---unfortunately, there is always a
polarimetric sacrifice to be made for any given design.  Linearly polarized
radiation (Stokes $Q$ and $U$)---especially weak polarization---is best
measured via the cross-correlation of the orthogonal outputs from native
circular feeds (with a penalty in measuring Stokes $V$); likewise,
circularly polarized radiation is best measured by cross-correlating the
outputs from orthogonal native linear feeds (with a penalty in measuring
Stokes $Q$).  New interferometers, except for the JVLA, are employing
native dual-linear feeds. While these feeds can be converted to
dual-circular by means of hybrids or dielectric quarter-wave phase
shifters, such devices provide frequency-dependent polarization, operate
over narrow bandwidths, and can increase not only the system temperature, but also
the complexity and the cost of the receiver design.

An altitude-azimuth (alt-az) mount is ideal for polarization measurements because it causes a
linearly polarized astronomical source to rotate relative to the feed as a
source is tracked. This rotation greatly reduces systematic effects and
makes calibration much easier. The Australian Square Kilometre Array Pathfinder (ASKAP) interferometer has been designed
with dishes on alt-az mounts, but the designers have opted to rotate their
entire dish structure with the sky---mimicking the behavior of an equatorial
mount---in order to simplify wide-field polarimetric imaging.  This is even better than a standard
alt-az mount because the polarization rotation can be done
instantaneously, without having to wait for the angle to change while tracking.

A final polarization consideration involves the choice to use an off-axis Gregorian secondary-focus design for the dishes. When it comes to measuring polarization, symmetry is better: sidelobes and spillover radiation tend to be polarized by an amount that increases with asymmetry. Moreover, ground radiation is highly polarized.

\section{Uniqueness to ngVLA Capabilities}

The centimeter-wave/$\sim$\,30\,GHz capabilities of the ngVLA are unique in that the other extant telescope that can currently perform such observations---namely, the JVLA---doesn't have the sensitivity to perform the type of observations we require. In the case of future ALMA Band 1 observations at $\sim$\,1\,cm wavelengths, the ngVLA's resolution and sensitivity will be much higher, enabling the studies we discuss above.  Lower frequency telescopes probe physical phenomena that are completely different from the various polarization mechanisms we discuss above.

\section{Synergies at Other Wavelengths}

\sloppy{
For continuum science, as an interferometer the ngVLA will fill a gap between the Square Kilometre Array (SKA)---which will address many different topics related to magnetic fields and polarization---and ALMA, which will look at many of the same objects, but at high frequencies where the material in the regions of interest (i.e., the inner parts of protoplanetary disks) is sometimes optically thick. The ngVLA will perform extremely high-resolution, high-sensitivity, targeted polarization science at the same time that the Origins Space Telescope (OST), if funded and constructed, is performing surveys of continuum and spectral-line polarization toward entire star-forming clouds across the Galaxy.  The ngVLA will perform long-wavelength follow-up of observations from the myriad telescopes that are currently available (or are soon coming online) and have polarization capabilities at far-infrared and (sub)millimeter wavelengths, including ALMA, SOFIA (HAWC+ polarimeter), the James Clark Maxwell Telescope (JCMT; POL2 polarimeter), the Large Millimeter Telescope (LMT; TolTEC polarimeter), and the BLAST-TNG balloon-borne experiment.
}

\bigskip
\acknowledgements  

The authors thank Carl Heiles for his help in clarifying the implications of the ngVLA receiver design on the telescope's polarimetric capabilities.
C.L.H.H acknowledges the support of the JSPS KAKENHI grant 18K13586.
J.M.G. is supported by the MINECO (Spain) AYA2017-84390-C2 grant.
R.G.-M. acknowledges support from UNAM-PAPIIT program IA102817.
P.K.G.W. acknowledges support for this work from the National Science Foundation through Grant AST-161477.
The National Radio Astronomy Observatory is a facility of the National Science Foundation operated under cooperative agreement by Associated Universities, Inc.

\bibliography{ms}
\bibliographystyle{asp2014}

%

\end{document}